\documentclass[aps,prl,twocolumn,letterpaper,superscriptaddress,showpacs]{revtex4-2}
\usepackage{graphicx}
\usepackage{CJK}
\usepackage{verbatim}
\usepackage{physics}
\usepackage[colorlinks,linkcolor=blue,anchorcolor=blue, citecolor=blue,urlcolor=blue,]{hyperref}
\usepackage{lineno}
\usepackage{bm}
\usepackage{mathptmx}
\usepackage{textcomp}
\usepackage{gensymb}
\makeatletter

\begin{document}
\begin{CJK*}{UTF8}{gbsn}
\title{Orbital dimerization-induced first-order structural phase transition: a case study in La$_3$Ni$_2$O$_7$}

\author{Xingchen Shen (沈星辰)}
\affiliation{School of Physics and Astronomy, Shanghai Jiao Tong University, Shanghai 200240, China}
\author{Wei Ku ({\CJKfamily{bsmi}顧威})}
\altaffiliation{corresponding email: weiku@sjtu.edu.cn}
\affiliation{School of Physics and Astronomy, Shanghai Jiao Tong University, Shanghai 200240, China}
\affiliation{Key Laboratory of Artificial Structures and Quantum Control (Ministry of Education), Shanghai 200240, China}
\affiliation{Shanghai Branch, Hefei National Laboratory, Shanghai 201315, China}

\date{\today}

\begin{abstract}
First-order structural phase transition is a common phenomenon in materials that qualitatively alters their physical properties.
Yet, the abrupt first-order nature
is usually unexplained by realistic computations, implying an omission of important physics in describing the electronic structure of the nearby stable phases.
Using the recently discovered nickelate superconductors La$_3$Ni$_2$O$_7$ as a prototypical example, we demonstrate that such first-order nature is typically beyond intra-atomic correlation considered in state-of-the-art material computations.
Instead, a full many-body treatment of low-energy active orbitals reveals a generic inter-atomic ``orbital dimerization'' mechanism of first-order structural phase transition, corresponding to \textit{abrupt} energy reduction upon a spin-singlet bond formation.
Such an inter-atomic correlation qualitatively changes not only the essential lattice bonding but also the characteristics of low-energy electronic properties across the transition.
This strong mechanism and the developed computational framework are generally applicable to a wide variety of ionic materials, to produce valuable insights into atomic and electronic structures essential for their physical properties and functionalities.
\end{abstract}

\maketitle
\end{CJK*}

First-order structural phase transitions are commonly observed in a wide variety of materials~\cite{st1,st2,st3,st4,st5,st6,st7,st8,st9,st10,st10b,st10c,st11,st12,st13} and often accompanied by \textit{qualitative} changes of important physical properties.
The most well-known example is the liquid-solid phase transitions, in which in addition to the frozen low-energy dynamics of atoms, even the electronic (optical and transport) properties qualitatively change, for example, from being metallic to insulating.
In functional materials, often even richer first-order solid-to-solid structural transitions are routinely observed.
For example, FeSe displays a first-order structural transition at high pressure, accompanied by a striped electronic antiferromagnetism that dictates the low-temperature physics~\cite{st10}.
Similarly, BaFe$_2$As$_2$ and SrFe$_2$As$_2$ undergo a first-order tetragonal-to-orthorhombic structural transition at low temperature, accompanied by an electronic antiferromagnetic order~\cite{st10b, st10c}.
These representative examples essentially reflect the fact that the occurrence of first-order phase transition affects not only the nature of lattice bonding but also the characteristics of the electronic structure that controls the electronic, magnetic, and superconducting properties of the functionality of materials.

This general implication of first-order phase structural transition manifests itself again in the recently discovered bilayer Ruddlesden-Popper nickelate La$_3$Ni$_2$O$_7$.
Under pressure, a first-order orthorhombic \textit{Amam} structure~\cite{struct} to the high-pressure \textit{Fmmm}~\cite{meng,jacs} or \textit{I4/mmm}~\cite{jacs,cxh} structural transition is observed.
The first-order nature is further confirmed by the coexistence of these phases even at the ambient pressure~\cite{cxh}.
Not too surprisingly, closely related to the first-order transition, experiments also found \textit{qualitative} changes of its electronic properties at high pressure, for example, from a Fermi liquid to a strange metal~\cite{meng,sm,dw1}, accompanied by suppression of a density wave-like order~\cite{dw1,dw2,dw3} and emergence of superconductivity~\cite{meng,sc1,sc2}.

Such dramatic first-order phase transitions must originate from some rather strong physical mechanism of relatively high energy.
This is evident from the puzzling high-pressure lattice structure in La$_3$Ni$_2$O$_7$.
Compared with the low-pressure structure, the high-pressure structure is characterized by the abrupt straightening of the tilted NiO$_6$ octahedra in the bilayer structure of the system~\cite{meng,jacs,cxh}.
Yet, counterintuitively, even with straightened NiO$_6$ octahedra, the interlayer distance is \textit{decreased} instead~\cite{cxh}.
This indicates that the underlying mechanism is energetically strong enough to qualitatively change the \textit{nature} of the eV-scaled atomic bonding.

Generally, such a strong mechanism is bound to \textit{abruptly} modify the electronic correlation and in turn the stable lattice structure, as in chemical bond formation and breaking~\cite{bbreak1,bbreak2}.
As demonstrated below, such a strong non-linear mechanism is beyond the capability of the \textit{smooth} enhancement of the kinetic hopping (even though this straightforward \textit{result} of the changed lattice is often regarded as an ``explanation''~\cite{ba1,ba2,ba3} for the distinct physical properties).
Therefore, in the presence of a first-order structural phase transition, identifying the underlying physical mechanism is essential not only for explaining the lattice structure but also for acquiring a proper understanding of the distinct electronic structures that control properties and functionalities of materials.

For such a commonly observed and essential phenomenon, physical understandings for first-order structural transitions are, however, mostly absent in the literature.
Even with the state-of-the-art computational studies~\cite{ma1,ma2,ma3,ma4,ma5,pt1,pt2,pt3,pt4}, in which different structures can stabilize under pressure or other structural constraints, the first-order nature of the structural transition is rarely reproduced or physically understood.
On the structural transition in La$_3$Ni$_2$O$_7$, for example, density functional theory (DFT)~\cite{dft1, dft2}-based calculations~\cite{meng,pt2} gave a total energy crossing for the \textit{Amam} and \textit{Fmmm} symmetry-constrained structures.
While this indicates different structures under various pressure, it does not address the essential first-order nature of the transition, let alone its understanding.

Here, we identify a generic ``orbital dimerization'' mechanism to address this long-standing general issue on proper understanding of first-order structural phase transitions in materials.
Specifically, using the recently discovered nickelate superconductor La$_3$Ni$_2$O$_7$ as a prototypical example, we show that such first-order nature is beyond the intra-atomic correlation considered in current state-of-the-art material computations.
Instead, a full many-body treatment of low-energy active orbitals finds an abrupt energy reduction due to \textit{inter}-atomic orbital dimerization through spin-singlet bonding.
Such an inter-atomic correlation also offers valuable insights into the distinct electronic properties at high pressure, such as the emergence of superconductivity in this material at high pressure.
This strong mechanism and the developed computational framework are generally applicable to ionic materials with open $d$- and $f$-shells, to provide proper understanding of both atomic and electronic structures essential for their physical properties and functionalities.

\begin{figure}[!t]
  \centering
  \includegraphics[width=0.95\linewidth]{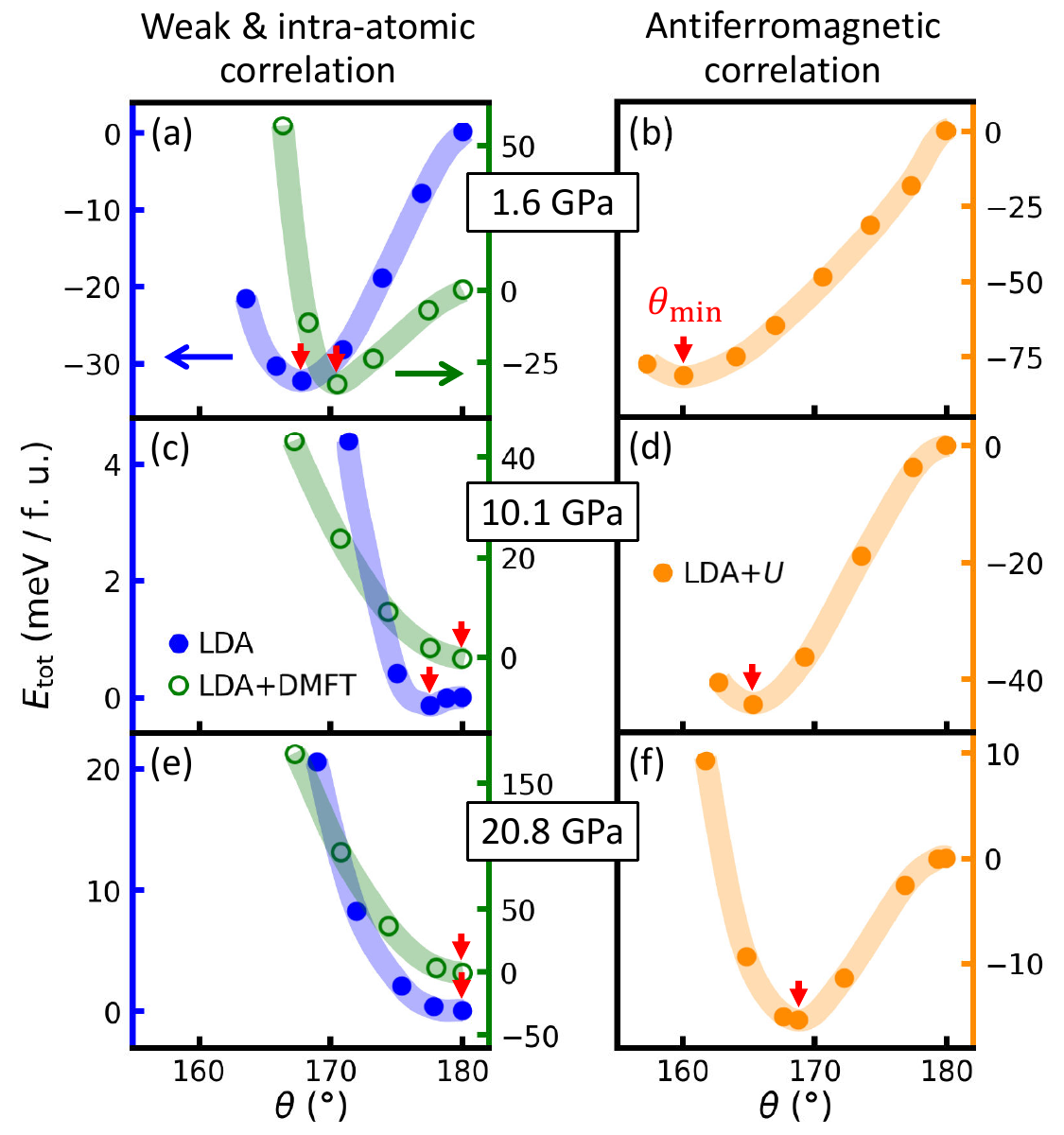}
  \vspace{-0.3cm}
  \caption{\textit{Lack of first-order phase transition in standard treatments} -
    System total energies against octahedral tile angle $\theta$ at low (a)(b), intermediate (c)(d), and high (e)(f) pressures, computed via standard LDA, LDA+DMFT, and LDA+$U$ treatments.
    The lattice structures for each constrained octahedral tile angle are optimized within each treatment with the same $U=6~{\rm eV},\ J=0.9~{\rm eV}$.
    The optimal tilt angels ($\theta_\mathrm{min}$, indicated by the red arrows) smoothly approach $180^\degree$ at higher pressure, displaying a \textit{continuous} structure transition in contradiction to experimental observations.}
  \label{fig:1}
  \vspace{-0.5cm}
\end{figure}

\textit{Insufficiency of standard treatments} --
Consider the recently discovered nickelate superconductor, La$_3$Ni$_2$O$_7$, as a prototypical example.
Using experimental lattice constants in the low-, intermediate-, and high-pressure phase, we start by performing the standard LDA calculations~\cite{dft2} for the system's total energy, with structures fully optimized~\cite{supplementary} under the constraint of various Ni-O-Ni angles, $\theta$.

The blue set of data in Fig.~\ref{fig:1}(a) shows that at low pressure the optimized structure hosts tilted Ni-O octahedra with $\theta_\mathrm{min}\sim 168\degree$, in good agreement with experimental observations~\cite{struct,meng,jacs,cxh} and previous LDA studies~\cite{meng,pt2}.
As the pressure increases, the optimal tilt angle \textit{continuously} reduces, for example to $\theta_\mathrm{min}\sim 178\degree$ at 10.1 GPa in Fig.~\ref{fig:1}(c).
Finally, at high pressure in Fig.~\ref{fig:1}(e), the octahedra straighten up as $\theta_\mathrm{min}$ smoothly reaches $180\degree$.
Clearly, even though the LDA treatment is able to reproduce the experimental ground-state lattice structures at low and high pressures, it completely misses the first-order nature of the transition.
Correspondingly, it fails to simultaneously stabilize both structures over the entire pressure range here, in direct contradiction to the recent experimental observation~\cite{cxh}.

Similar smooth behavior is found upon \textit{fully} incorporating intra-atomic correlation through the state-of-the-art LDA+DMFT method~\cite{dmf1,dmf2}.
Indeed, the green set of results of LDA+DMFT in Fig.~\ref{fig:1}(a)(c)(e) display a similar trend as those of LDA, only with a slightly weaker tendency toward octahedral tilt and thus a $\theta_\mathrm{min}$ closer to $180\degree$.
Therefore, the essential physics for the first-order nature of the structural transition must require \textit{inter}-atomic correlations.

\textit{Incorporating inter-atomic correlation} --
A simple way to simulate such an inter-atomic correlation is via introduction of antiferromagnetic correlation in the spin \textit{density} using the LDA+$U$ method~\cite{pU1,pU2}.
However, Fig.~\ref{fig:1}(b)(d)(f) show that such simulation suffers from the well-known under-binding tendency~\cite{underbind1,underbind2} of Hartree-Fock treatment of interaction in LDA+$U$, such that the octahedra remain very tilted even at the pressure as high as 20.8GPa.
This rather negative result indicates that \textit{full} many-body treatment of inter-atomic correlations (beyond the LDA+$U$ reproduction of spin density correlation), particularly their \textit{energetic} contributions, is essential to a proper description of the lattice bonding.

To this end, we introduce a simple improvement to the density functional theory (DFT) computation of total energy, by incorporating a full many-body treatment of the effective Hamiltonian,
\begin{equation}
    H=T+V-\langle T+V\rangle_{\rm dc}+E_{\rm DFT},
\label{H_eff}
\end{equation}
containing one-body kinetic and potential processes, $T$, and intra-atomic repulsion, $V$, of only the most correlated subspace of low-energy orbitals.
Specifically,
\begin{equation}
    V=
    \frac{1}{2}\sum_{\substack{j,\{m\}\\\sigma\sigma'}}
    \langle m_1,m_2|V_{ee}|m_3,m_4\rangle 
    c^\dagger_{\substack{jm_1\sigma}}
    c^\dagger_{\substack{jm_2\sigma'}}
    c_{\substack{jm_4\sigma'}}
    c_{\substack{jm_3\sigma}},
\end{equation}
between \textit{atomic} Wannier orbitals~\cite{wan}, $c^\dagger_{\substack{jm\sigma}}$, of index $m$ and spin $\sigma$ at atomic site $j$ is approximated via the standard Slater integrals, $V_{ee}$, as parameterized in the LDA+$U$ and LDA+DMFT methods.
The one-body processes in $T$,
\begin{equation}
    T=
    \sum_{\substack{jj^\prime mm^\prime}\sigma}
    t_{j m, j^\prime m^\prime}
    c^\dagger_{\substack{jm\sigma}}
    c_{\substack{j^\prime m^\prime \sigma}}
\label{t_eff}
\end{equation}
are obtained from the Wannier representation~\cite{wan,lee,Jrs} of the self-consistent LDA+$U$ Kohn-Sham Hamiltonian, with the interaction-induced orbital energy removed through the atomic limit formula~\cite{pU2}.
The improved total energy of the system, $\langle H\rangle$, is then simply the eigenenergy of the \textit{many-body} ground state, as evaluated via exact diagonalization of $H$.

Similar to the LDA+$U$ and LDA+DMFT approaches, Eq.~\ref{H_eff} needs to estimate the approximate energy contribution of $T+V$ already included in the LDA+$U$ total energy, $E_\mathrm{DFT}$, and remove such double counting (dc), $\langle T+V\rangle_{\rm dc}$.
Since LDA+$U$ treats the intra-atomic interaction via Hartree-Fock approximation, $\langle T+V\rangle_{\rm dc}$ is therefore consistently estimated via the self-consistent Hartree-Fock treatment of $T+V$ within the subspace of active orbitals.
This way, the energy improvement, $\Delta E\equiv \langle H \rangle - E_\mathrm{DFT}=\langle T+V\rangle-\langle T+V\rangle_{\rm dc}$, would contain only the many-body correlation energy beyond the energy of the intra-atomic Hartree-Fock states.
($\Delta E\to 0$ if the many-body ground state is restricted to one with optimal single-determinant wavefunction, as in the LDA+$U$ approximation.)

Note that $E_\mathrm{DFT}$ and $T$ are obtained directly from DFT results, so they need to be computed for \textit{each} atomic lattice to incorporate all the higher-energy electronic and ionic  contributions to the DFT total energy.
Also, the spatial density distribution of the many-body ground state can, in principle, differ from that of the DFT results, so the above procedure can be extended to account for the density self-consistence with the DFT framework.
Nonetheless, given the rather large energy scale of the intra-atomic Coulomb energy and that of the inter-atomic charge transfer with the ligands, in most cases DFT density is already very robust.
Instead, the improvement from the full many-body treatment is mainly in the additional energy reduction associated with many-body fluctuation (which is poorly approximated in common DFT energy functionals).
Particularly, since this study aims at illustrating a strong physical mechanism that gives rise to an abrupt total energy reduction, one expects such a dominant effect to reveal itself regardless of the minor improvement in density.

For the specific case study of La$_3$Ni$_2$O$_7$, previous experimental~\cite{s7,s8,s9,s10} and theoretical~\cite{jiang,th1,th2,th3} studies have established a strong charge-transfer nature of the system.
Therefore, the most relevant low-energy Hilbert space consists of two subspaces whose charge dynamics are nearly decoupled from each other by their distinct $z$-parity~\cite{jiang}: 1) the in-plane O atoms hosting ligand holes that fluctuate to the Ni $d_{x^2-y^2}$ orbitals, and 2) the Ni $d_{z^2}$ orbitals hosting spin dynamics mediated by the apical O atoms between the layers.
Since the abrupt change of the lattice structure associated with the first-order transition is on the interlayer Ni-O-Ni bond, the most essential many-body Hilbert space thus includes the full degrees of freedom of $d_{eg}$ orbitals of Ni atoms in both layers and the $p_z$ orbitals of the apical O atom in between.

Conveniently, the most recent experimental structural data~\cite{cxh} found both stable structures with tilted and straightened octahedra over the entire pressure range.
This provides the essential lattice parameters for us to fully optimize the electronic and lattice structure at both ambient pressure and about 16 GPa within DFT, under a \textit{series} of constrained Ni-O-Ni bond angles, $\theta$.
The resulting DFT solution is then used to construct the effective $H$ in Eq.~\ref{H_eff} containing the most relevant many-body subspace mentioned above.
The improved ground state energy, $\langle H \rangle$, is then obtained via an exact diagonalization of $H$ for each constrained $\theta$.

\begin{figure}[!t]
  \centering
  \includegraphics[width=0.95\linewidth]{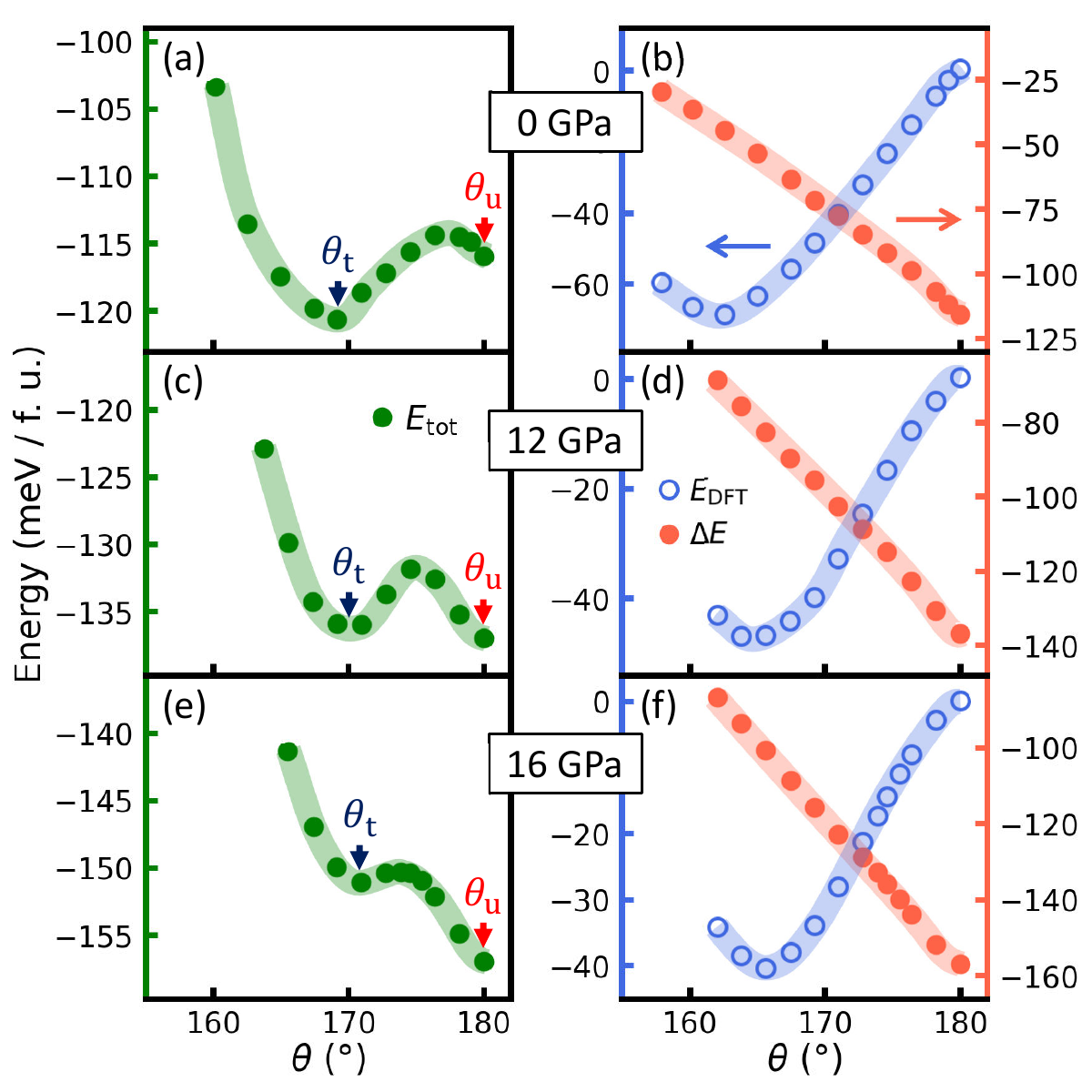}
  \vspace{-0.4cm}
  \caption{\textit{First-order phase transition via inter-atomic orbital dimerization} -
    (a)(c)(e) System total energy with inter-atomic correlation display two stable structures with a tilted ($\theta_\mathrm{t}\sim 170^\degree$) and untilded ($\theta_\mathrm{u}=180^\degree$) octahedra.
    Around 12 GPa, the energy minimum \textit{jumps} from the tildted octahedra to the untilted ones, indicating a first-order structural transition in agreement with experiments.
    (b)(d)(f) The essential double-minimum structures originate from a strong energy reduction, $\Delta E$, from the LDA+$U$ energy, $E_\mathrm{DFT}$, near $\theta\sim 180^\degree$ due to inter-atomic orbital dimerization associated with an enhanced spin-singlet bond.}
  \label{fig:2}
  \vspace{-0.5cm}
\end{figure}

\textit{First-order nature of structural transition} --
Let's consider first the high-pressure case shown in Fig.~\ref{fig:2}(e).
In great contrast to the LDA+$U$ treatments, our resulting total energy of the system, $E_\mathrm{tot}=\langle H \rangle$, has a global minimum at untilted $\theta_\mathrm{u}=180^\degree$, nicely reproducing the experimental observation at high pressure~\cite{meng,jacs}.
More importantly, even at high pressure, the total energy shows another local minimum at $\theta_\mathrm{t}\sim 170^\degree$.
This indicates the local stability of the structure with tilted octahedra, in agreement with the experimental observation of a second stable structure~\cite{cxh}.
Such coexistence of two locally stable structures is precisely the signature of the experimentally observed first-order phase transition.

Similarly, at ambient pressure, Fig.~\ref{fig:2}(a) shows a similar double-minimum structure, again in agreement with the experimental observation~\cite{meng,jacs}.
Nonetheless, the structure with untilted octahedra at $\theta_\mathrm{u}$ remains a locally stable structure even at ambient pressure.
Such coexistence of two stable structures nicely reproduces the experimental observation at ambient pressure~\cite{cxh}.

Naturally, in the intermediate pressure the \textit{energies} of these two minima would smoothly approach each other.
Nonetheless, in great contrast to smooth increase of $\theta_\mathrm{min}$ under pressure in the LDA and LDA+DMFT results in Fig.~\ref{fig:1}, $\theta_\mathrm{t}$ of the tilted structures in Fig.~\ref{fig:2} remain around $170^\degree$ within this entire pressure range, in excellent agreement with the recent experiment~\cite{cxh}.
Thus, as the applied pressure increases to the critical pressure $P_c\approx12$ GPa [c.f. Fig~\ref{fig:2}(c)], the system's global energy minimum abruptly jumps from $\theta_\mathrm{t}$ to $\theta_\mathrm{u}$, thus displaying a first-order structural transition in perfect agreement with experiments~\cite{meng,jacs}.

\textit{General ``orbital dimerization'' mechanism} --
Theoretically, the essential ingredient for such a first-order phase transition is a strong energy reduction in the vicinity of $\theta\sim\theta_\mathrm{u}$ that produces an additional local minimum in $E_\mathrm{tot}$.
This is more clearly demonstrated by the strong energy improvement, $\Delta E\equiv \langle H \rangle - E_\mathrm{DFT}$, around $\theta\approx180^\degree$ in Fig.~\ref{fig:2}(b)(d)(f) via red solid circles.
Furthermore, away from $\theta\sim\theta_\mathrm{u}$, this energy reduction quickly weakens such that the other minimum from $E_\mathrm{DFT}$ (blue solid circles) remains stable.
This indicates the emergence of a \textit{new physical mechanism} near $\theta=180^\degree$ that drastically enhances the Ni-O-Ni interlayer bonding.

\begin{figure}[!t]
  \centering
  \includegraphics[width=0.95\linewidth]{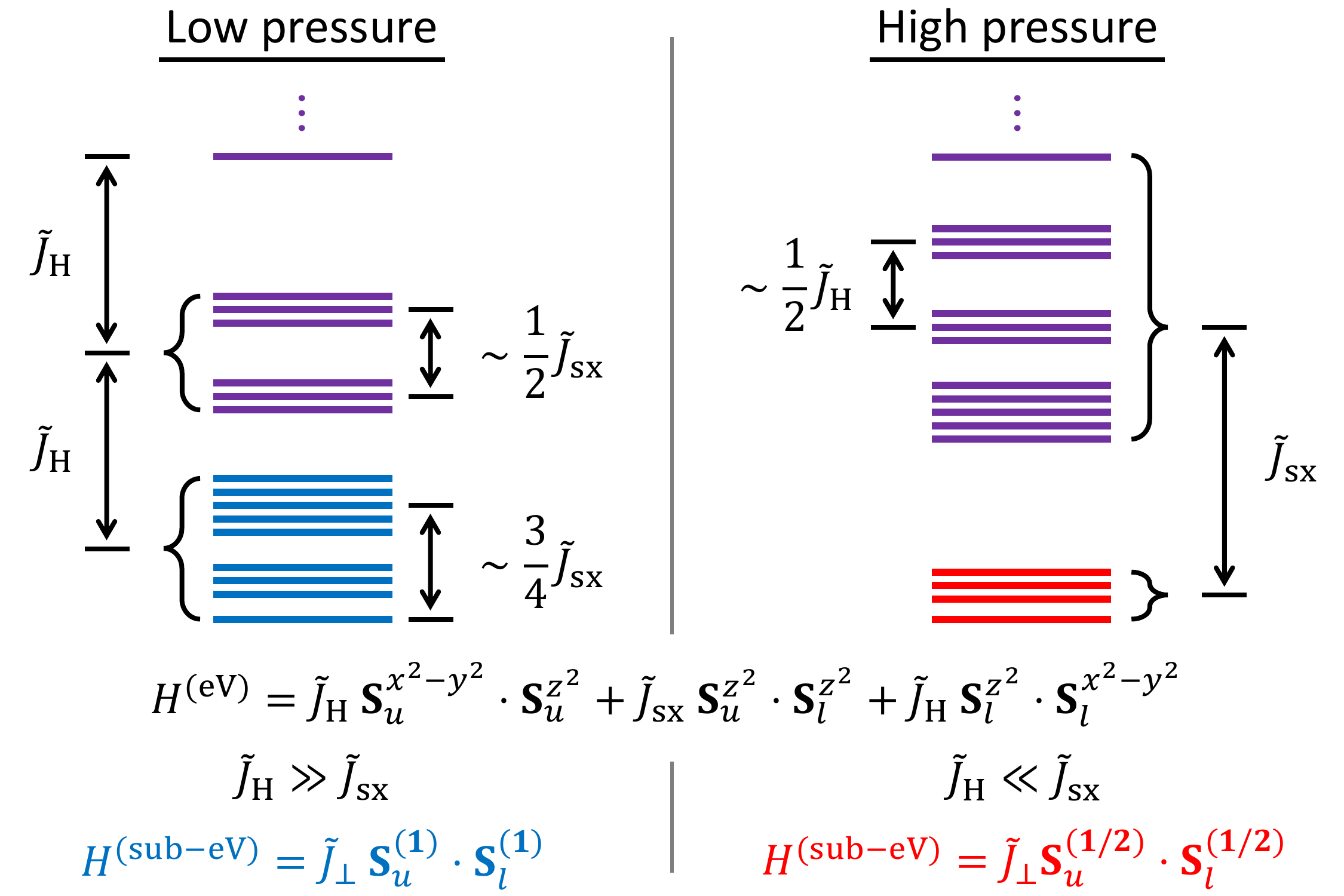}
  \vspace{-0.2cm}
  \caption{\textit{Distinct low-energy electronic structures} -
    The eV-scale eigen-energy structures of $H$ correspond to dynamics of 4 dressed 1/2-spins in the upper (u) and lower (l) layers with competing couplings.
    The low-pressure case is dominated by the renormalized intra-atomic Hund's coupling $\tilde{J}_\mathrm{H}$ that results in typical spin-1 Ni$^{2+}$ ions in the sub-eV dynamics (in blue).
    In contrast, the high-pressure case is dominated by the inter-atomic antiferromagnetic $\mathbf{S}^{z^2}_u$-$\mathbf{S}^{z^2}_l$ superexchange that effectively fractionalizes the Ni$^{2+}$ ionic spins from 1 to 1/2 in the sub-eV scale (in red).}
  \label{fig:3}
  \vspace{-0.5cm}
\end{figure}

The underlying physical mechanism can be deciphered through identifying the key difference in the resulting eigenstates and their eigen-energy structures at the two minima.
Specifically, as illustrated in Fig.~\ref{fig:3} the resulting eV-scale  eigenstates of $H$ correspond to the dynamics of 4 dressed 1/2-spins across the bi-layer with two main couplings (as previously found in a multi-energy analysis~\cite{jiang}): 1) renormalized intra-atomic Hund's coupling $\tilde{J}_\mathrm{H} \mathbf{S}^{z^2} \cdot \mathbf{S}^{x^2-y^2}$, and 2) emergent inter-atomic antiferromagnetic superexchange, $\tilde{J}_\mathrm{sx} \mathbf{S}^{z^2}_u \cdot \mathbf{S}^{z^2}_l$, between the Ni-$d_{z^2}$ orbitals in the upper ($u$) and lower ($l$) layers.
Compared with the tilted cases of the same pressure, one finds a systematic strong enhancement of the resulting $\tilde{J}_\mathrm{sx}$ in the untilted cases over the studied 20 GPa pressure range~\cite{supplementary}.
Evidently, the significant energy reduction near $\theta=180^\degree$ is associated with the emergence of this strong inter-layer superexchange and the corresponding spin-singlet correlation.

Essentially, these results point to a general ``orbital dimerization'' mechanism for enhancing covalent bonding beyond the typical one-body kinetics.
Recall that spin-singlet correlation between two 1/2-spins is one of the most efficient many-body effects in significantly lowering energy, owing to the off-diagonal quantum fluctuation, $\frac{1}{2}\tilde{J}_\mathrm{sx}(S^+_u S^-_l + S^-_u S^+_l)$.
When active, it would strongly promote dimerization of the associated \textit{orbitals} in neighboring atoms, similar to the bond formation in H$_2$ gas.
One thus expects such orbital dimerization to give an essential contribution to general covalent bonds in materials.
Especially in systems displaying structural transitions, this strong mechanism would naturally promote a first-order structural transition, when its corresponding bonding energy exceeds the inter-ionic Coulomb interaction (which instead prefers atomic close-packing as in the tilted structure.)

\textit{Distinct quantum states of matter across the first-order transition} --
As emphasized in the introduction, the first-order nature of the transition typically implies a rather strong physical mechanism that can qualitatively alters not only the atomic lattice structure but also the lower-energy \textit{electronic} structure.
Indeed, for the specific case of La$_3$Ni$_2$O$_7$, near $\theta=180^\degree$ the strong interlayer superexchange, $\tilde{J}_\mathrm{sx}$, effective binds the Ni-$d_{z^2}$ orbitals into \textit{inert} spin-singlet pairs and in turn removes them from the lower-energy dynamics of the ligand holes residing in the in-plane O ions.

Particularly, at high pressure when the kinetic hopping between $d_{z^2}$ and $p_z$ orbitals exceeds 2 eV, the resulting inter-atomic singlet-triplet splitting, $\tilde{J}_\mathrm{sx}$, quickly grows stronger to eV-scale and surpasses the renormalized intra-atomic Hund's splitting, $\tilde{J}_\mathrm{H}$.
Correspondingly, as illustrated in the right column of Fig.~\ref{fig:3}, the sub-eV structure of the many-body eigen-energies of $H$ resembles that of two 1/2-spins, rather than that of two spins-1 Ni$^{2+}$ ions found in the low pressure phase.
That is, under high enough pressure the Ni$^{2+}$ ionic spins are \textit{effectively} fractionalized from 1 into 1/2 in their low-energy dynamics, as found in the previous multi-energy analysis~\cite{jiang}.
This leaves only half-filled effective $d_{x^2-y^2}$ orbital active to magnetically couple to the in-plane ligand holes~\cite{jiang}, in astonishing resemblance to the sub-eV Hamiltonian of the cuprates~\cite{RVB,emery,zhang-rice,t-j}.
Therefore, this material's resemblance to the cuprates in non-Fermi liquid transport~\cite{sm} and unconventional superconductivity~\cite{meng} is perfectly understandable.

To the best of our knowledge, this orbital dimerization with the associated ionic spin fractionalization is the only mechanism to date that \textit{qualitatively} alters the high-pressure superconducting phase of this material from the low-pressure non-superconducting one.
In contrast, in most existing studies~\cite{ba1,ba2,ba3}, the change of parameters in the high pressure can only produce \textit{quantitative} modification in the low-energy physics and thus misses the essential qualitative differences due to such high-energy correlation, under which the observed superconductivity emerges.

Now beyond this specific case, given the very strong binding associated with singlet correlation, one expects the above orbital dimerization mechanism to be a rather common phenomenon, particularly in materials displaying pressure-induced first-order structural transitions.
Possible candidates include, for example, the iridates (Sr$_2$IrO$_4$ and Sr$_3$Ir$_2$O$_7$)~\cite{st9,st10}, whose pressure-induced first-order structural transition is accompanied by a sudden collapse of the interlayer lattice parameter, consistent with orbital dimerization.
Again, we stress that the presence of such first-order structural transition implies a high-energy correlation that likely alters qualitatively the lower-energy electronic structure as well.

\textit{Limitation of state-of-the-art computational methods} --
It is worth noting that the intrinsic \textit{two}-body nature of such \textit{inter}-atomic orbital dimerization renders this important correlation challenging to incorporate within the state-of-the-art computational methods.
Indeed, current standard material computations~\cite{dft1,dft2}, such as LDA, LDA+$U$, or LDA+DMFT, only explicitly incorporate at best the intra-atomic correlation, leaving the essential inter-atomic covalent bonding accounted for only by the effective one-body kinetics.
As demonstrated above, such approximations cannot properly incorporate the significant energy reduction associated with the essential quantum fluctuation of spin-singlet correlations.
They thus lack the necessary non-linearity in the system energy to properly reproduce the first-order structural transition.
More seriously, the omission of such strong correlation renders the low-energy electronic structure \textit{unreliable}.
This limitation, in essence, is similar to the well-known problem in describing the electronic structure and its abrupt energy change during typical molecular dissociation processes~\cite{disso1,disso2,disso3}.

Similarly, this type of strong \textit{non-perturbable} short-range correlations are beyond diagrammatic approximations, such as the \textit{GW}~\cite{GW1,GW2} or parquet~\cite{parquet} approximations.
In principle, more extended cluster approximation~\cite{CCA} or dynamical cluster approximation~\cite{DCA} can capture this effect, but their large computational expense has limited their application in material computations.
The many-body approach presented here thus offers a simplest solution to incorporate the leading energy correction from such inter-atomic correlation.

\textit{Conclusion} --
In short, we identify a general ``orbital dimerization'' mechanism for pressure-induced first-order structural phase transition, using the unconventional superconducting La$_3$Ni$_2$O$_7$ as a prototypical example.
Without explicitly incorporating inter-atomic correlation, the state-of-the-art theoretical treatments are shown to be insufficient in producing the first-order structural transition, let alone the qualitative change in the low-energy physics.
To this end, we develop a many-body framework to improve the system's total energy of density functional computations.
Upon including the full many-body effects of the low-energy active orbitals, the first-order structural transition is reproduced through the development of a strong interlayer singlet bond via an eV-scale superexchange.
The emergence of this additional bonding \textit{qualitatively} alters the lower-energy physics of the material, effectively fractionalizing the Ni$^{2+}$ ionic spin to 1/2 and producing an effective sub-eV dynamics similar to that of the high-temperature cuprates.
The identified orbital dimerization mechanism, together with our developed computational framework, is expected to apply to a wide range of functional materials, especially those displaying pressure-induced first-order structural transitions.

\begin{acknowledgments}
The authors acknowledge helpful discussions with Ruoshi Jiang. This work is supported by National Natural Science Foundation of China (NSFC) \#12274287 and \#12042507 and Innovation Program for Quantum Science and Technology 2021ZD0301900.
\end{acknowledgments}

\bibliography{MainTex.bib}

\end{document}